# The speed of sound and elastic properties of single crystals of inorganic and hybrid lead-free iodide perovskites obtained by femtosecond transient optical reflectivity

Giuseppe Ammirati[1], Patrick O'Keeffe[2], Stefano Turchini[1], Daniele Catone[1], Alessandra Paladini[2], Francesco Toschi[1], Stevan. Gavranovic[3], Jan Pospisil[3], Giovanni Mannino[4], Salvatore Valastro[4], and Faustino Martelli[1,5*]

[1] *Istituto di Struttura della Materia - CNR (ISM-CNR), EuroFEL Support Laboratory (EFSL), Via del Fosso del Cavaliere 100, 00133, Rome, Italy.*
[2] *Istituto di Struttura della Materia - CNR (ISM-CNR), EuroFEL Support Laboratory (EFSL), 00015Monterotondo Scalo, Italy.*
[3] *Brno University of Technology Brno 612 00, Czech Republic*
[4] *Institute for Microelectronics and Microsystems, (IMM-CNR), 95121 Catania, Italy*
[5] *Institute for Microelectronics and Microsystems, (IMM-CNR), 00133 Rome, Italy*

*email: faustino.martelli@cnr.it

**Abstract**

Optoelectronic devices operate under continuous thermal stress that may influence both long-term stability and optical properties of the active material. It is therefore useful to know the elastic properties of the active materials. In general, the elastic constants of a material may be deduced by the measurement of the speed of sound in that material. In this work, we report on the measurements of the speed of sound in three lead-free halide perovskites, namely the inorganic $Cs_3Bi_2I_9$ and the hybrid $MA_3Bi_2I_9$ and $MA_3Sb_2I_9$, by means of femtosecond transient optical reflectivity that shows oscillations of the signal intensity that are related to the strain field caused by the photoexcitation of electron-hole pairs. The values of the speed of sound thus obtained have allowed us to extract the elastic constant, $c_{11}$, along the propagation direction of the light. The $c_{11}$ values indicate a lower stiffness of the hybrid materials, an important aspect when designing optoelectronic devices.

Optoelectronic devices are exposed to thermal and mechanical stress because of the alternate on/off status or in the packaging process or, as in the case of solar cells, because of the alternation of day and night. The different types of stress may influence the long-term stability and optical properties of the active material. On the other hand, it is possible to design devices that exploit elastic deformation to improve the device performance as it has been proposed, e.g., to modulate the optical and electronic properties of photovoltaic materials [1]. It is therefore very important and useful to know the elastic properties of the materials that make up optoelectronic devices. In general, the elastic constants of a material may be deduced by the measurement of the speed of sound via the relationship:

$$V = \sqrt{\frac{c_{ij}}{\rho}} \quad (1)$$

Where $V$ is the speed of sound, $c_{ij}$ is the elastic tensor ($i$ and $j$ describe the direction of stress and deformation, respectively), describing the stiffness of the material in response to a stress and $\rho$ is the material density. In crystals, the elastic properties are generally anisotropic.

During the last decade, halide perovskites (HPs) have emerged as active materials in solar cells allowing for highly efficient photovoltaic conversion with a certified value of 26.7%. [2] In addition to the obvious potential for their use in real devices, lead-based HPs present some problems to be solved, including the presence of lead in their composition, due to the toxicity of this material. Research on alternative, environmentally friendly elements that may replace $Pb^{2+}$ in the crystalline lattice of HPs for use in solar cells has therefore become an important part of the studies carried out on this class of materials [3–6]. Among them, bismuth and antimony are promising candidates [7–12]. They emerged as high-profile semiconductor material, finding applications not only in photovoltaics but also for x-ray detectors [13], and photodetectors [14].

In these frameworks, we have measured the speed of sound for three lead-free, inorganic or hybrid, Bi- and Sb-based HPs, using femtosecond transient reflectivity (FTR) and exploiting the generation of coherent acoustic phonons (CAPs) [15], namely ultrasonic strain pulses with frequencies ranging from GHz to THz, through the photoexcitation of carriers in semiconductors by ultrafast light pulses. The absorption of a laser pulse with a short optical penetration depth causes sudden expansion deformation of the lattice (thermal stress) [16]. The generation of CAPs is equivalent to the creation of an elastic strain field that perturbs the light reflection in the semiconductor via the change of the complex refractive index. [17] The generation of a strain pulse can be modeled in terms of the deformation potential coupling with photoexcited carriers, and its detection in terms of the photoelastic effect. [18] The interaction of light with coherent acoustic phonons has been studied in various semiconductors for different scopes [16,18–24], including hybrid lead-HPs like $CH_3NH_3PbI_3$ [19] and inorganic lead-free perovskites like $Cs_2AgBiBr_6$ [25] and $Cs_3Bi_2I_9$. [24]

In this work, we investigate the intrinsic generation and detection of coherent acoustic phonons in single crystals (SCs) of the indirect band-gap iodide perovskites $Cs_3Bi_2I_9$, $MA_3Bi_2I_9$, and $MA_3Sb_2I_9$. Using FTR, we were able to extract the speed of sound and the $c_{11}$ elastic constant for these three materials, providing new insights into the properties of these lead-free perovskites and to point out differences due to the presence of an inorganic (Cs) or organic (MA) cation in the material composition.

In pump-probe FTR measurements, the elastic field caused by the generation of coherent acoustic phonons by the pump pulse influences the propagation of the probe light. In our set-up, the pump and probe propagate in the same direction thus allowing the measurement of the sound speed in that same direction and the estimate of the $c_{11}$ elastic constant. In the pump-probe experiment, the probe light intensity will show sinusoidal temporal dynamics, known as Brillouin oscillations, whose period, in the case of samples thicker than the light penetration depth, as in the case of our SCs, decreases with decreasing probe wavelength and with increasing light penetration depth. In the limit of propagation length shorter than sample thickness and normal incidence, the oscillation period is given by the following formula [18]:

$$\frac{1}{\tau} = \frac{2n(\lambda)v}{\lambda} \qquad (2)$$

where n is the refractive index, and $\lambda$ is the probe wavelength in air. We anticipate that the oscillation intensity does not decay in the temporal window of our experiment (-1 to 300 ps) because the acoustic pulse acts on the probe propagation up to its penetration depth, which is expected to be in the micrometer range because of the indirect character of the bandgap of the investigated materials [26].

Our iodide-based perovskite SCs were grown by hydrothermal methods [27]. The detailed synthetic procedure and microscopic images of the prepared single crystals are shown in **Section S1** of the Supplemental Material (SM). The thickness of the SCs was of several hundred µm.

The FTR setup uses an amplified femtosecond laser pulse with a duration of 35 fs, repetition rate of 1 kHz, and a power output of 4 W, centered at 800 nm. Pump pulses at 4.5 eV were generated through an optical parametric amplifier. The probe consists of a white-light supercontinuum beam spanning 360–780 nm (1.60–3.45 eV), which is produced by directing 3 µJ of the 800 nm pulse into a rotating $CaF_2$ crystal. The pump and probe beams are focused onto the sample with spot diameters of 200 µm and 150 µm, respectively. The delay time between the two beams is adjusted by varying the optical path length of the probe, yielding an instrument response function of approximately 50 fs. Further details about the experimental setup are available in previous studies.[26,27] The pump fluence was 600 µJ/cm$^2$. A variation of the pump fluence only leads to a linear variation of the signal intensity, with no impact on the physical meaning of the spectra. The incidence angle of both light beams was approximately 7° with respect to the normal incidence on the (001) plane of the single crystals.

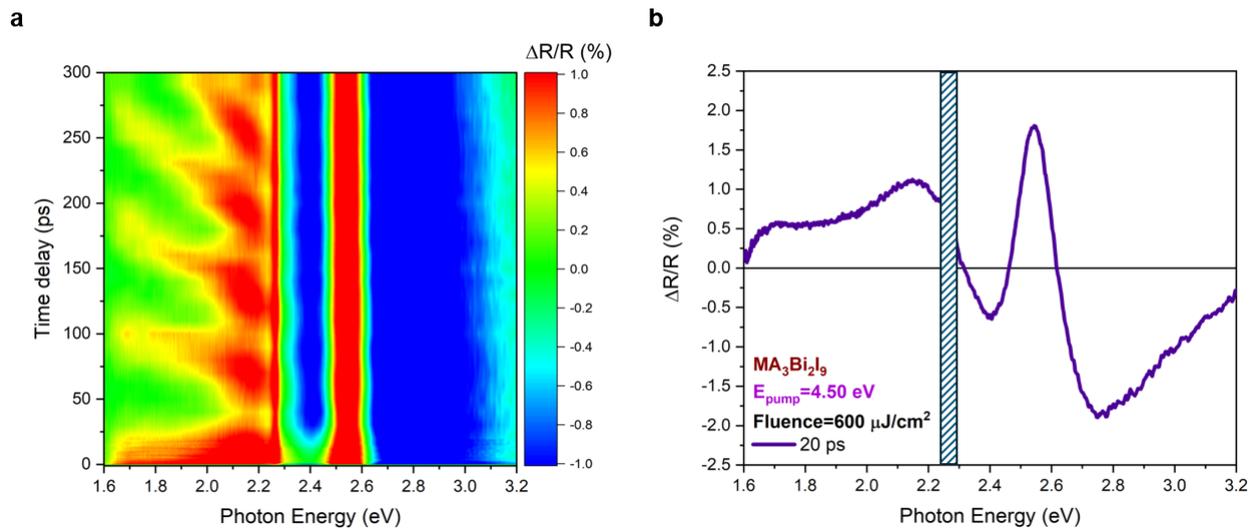

**Figure 1:** (a) False color TR map obtained for MA$_3$Bi$_2$I$_9$ for pump energy of 4.50 eV and fluence of 600 μJ/cm$^2$. (b) TR spectrum obtained at the time delay of 20 ps. The positive signals indicate absorption bleaching while the vertical dashed band covers the spectral range where the scattered pump light dominates over the TR signal.

**Figure 1** shows the false-color transient reflection (TR) map (a) for MA$_3$Bi$_2$I$_9$, along with the TR spectrum, taken at a pump-probe time delay of 20 ps. An analogous figure is reported in **Section S2** of the SM for the other two materials investigated here, the inorganic Cs$_3$Bi$_2$I$_9$ and the hybrid MA$_3$Sb$_2$I$_9$, (see **Figure S4**). The excitonic energy (about 2.6 eV) found with FTR is in very good agreement with the results obtained by spectroscopic ellipsometry and the Elliot fit reported by Valastro et al. [27]. This agreement confirms that the FTR measurements reflect the properties of the materials under investigation.

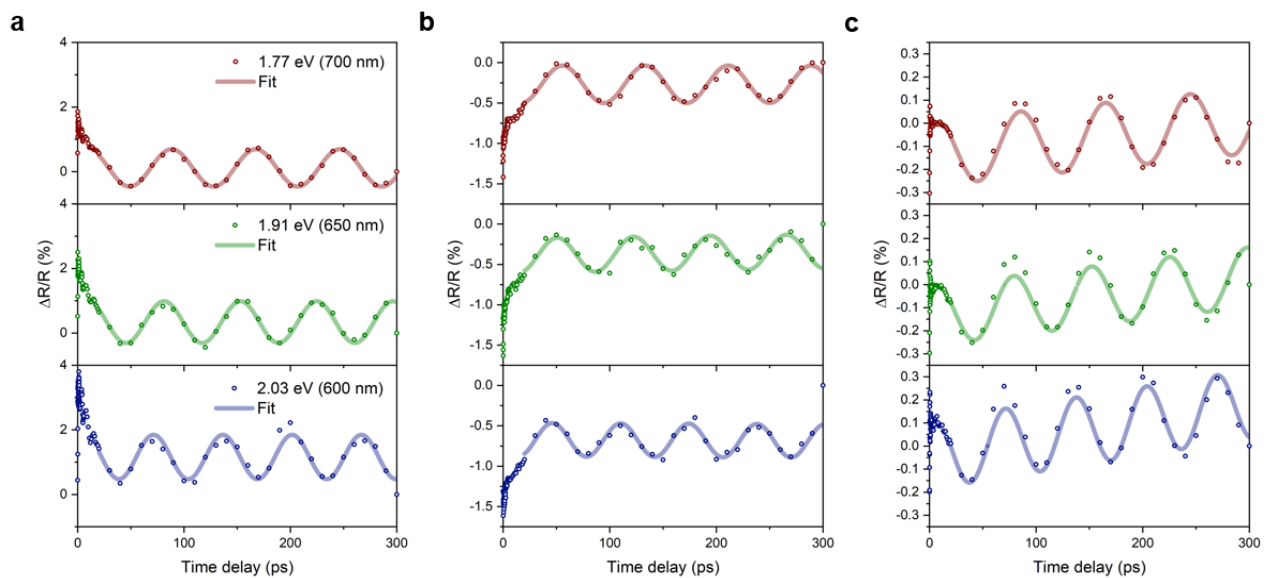

**Figure 2:** Temporal dynamics of the transient reflectivity signal in $Cs_3Bi_2I_9$ (a), $MA_3Bi_2I_9$ (b), and $MA_3Sb_2I_9$ (c) obtained at the probe photon energy of 1.77 eV (red points and line), 1.91 eV (green), and 2.03 eV (blue) and their relative fitting curves, as explained in the main text. The TR spectra have been measured with pump photon energy of 4.50 eV and fluence of 600 µJ/cm²

We will not discuss here the spectra details and dynamics of the excitonic absorption bleaching, which agree with those found in the literature [30], as in this work our interest is mainly devoted to the elastic properties of the materials. Our attention goes to the almost periodic features that are observed in the FTR, for energies below the excitonic resonance, see **Figure 1a**, where no spectral feature related to the band structure is observed. We report in **Figure 2** the temporal dynamics of the TR intensity for three different probe energies (1.77, 1.91, and 2.03 eV) for all three materials. The intensity of the FTR spectra shows periodic oscillations with a period of tens of picoseconds. These oscillations appear at a time delay of 20 ps and are not related to the dynamics of the carriers photoexcited by the pump and persist for hundreds of picoseconds. As mentioned in the introduction, they are due to the modulation of the probe light propagation into the material by the strain field induced by the coherent acoustic phonons generated by the thermal stress originated by the carriers photoexcited by the pump [16]. This modulation moves with the velocity (*V*) of the longitudinal acoustic phonons. The use of a broad range of probe energies facilitates the validation of observed phenomena and allows the investigation of the dependence on the probe energy, ensuring consistency and reducing the likelihood of artifacts arising from wavelength-specific effects.

The intensity of the oscillations is fitted as follows:

$$\Delta A = (a_1 t + a_0) + A \cdot \cos\left(\frac{2\pi}{\tau}(t - t_0) + \theta\right) \tag{3}$$

where ($a_1 t + a_0$) is a straight line fitting the background lying beneath the oscillations, A is the amplitude of the oscillation, τ is the oscillation period, θ is the phase, and $t_0$ is the initial time of the oscillation (20 ps).

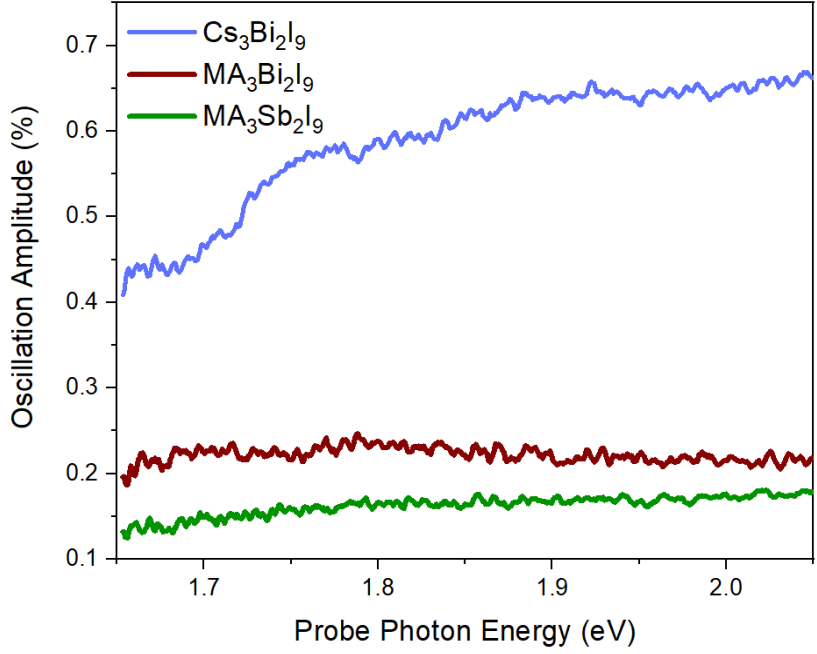

**Figure 3:** The amplitude of the oscillations of the transient reflectivity as a function of the probe photon energy for $Cs_3Bi_2I_9$ (blue line), $MA_3Bi_2I_9$ (red line) and $MA_3Sb_2I_9$ (green line). The pump photon energy is 4.50 eV and the fluence 600 µJ/cm².

In **Figure 3**, we show the dependence of the oscillation amplitude on the probe energy. The amplitude is larger in the inorganic $Cs_3Bi_2I_9$ than in the two hybrid compounds. $Cs_3Bi_2I_9$ also shows a different probe energy dependence with respect to the hybrid materials. As shown in the work by Valastro and coworkers [27], the absorption coefficient of $Cs_3Bi_2I_9$ is very similar to that of $MA_3Bi_2I_9$. In particular, the value at the pump energy (4.50 eV) differs by less than 10%. Therefore, the pump excites a similar density of e-h pairs in the two materials giving rise to similar changes in the complex refractive index. The observed difference in the amplitude of the oscillations should be attributed to a different deformation potential of the energy bands as the strain occurs in the two materials [31], that reflects differences in the elastic constants (see below). On the other hand, the probe energy dependence cannot be explained in simple ways. [18]

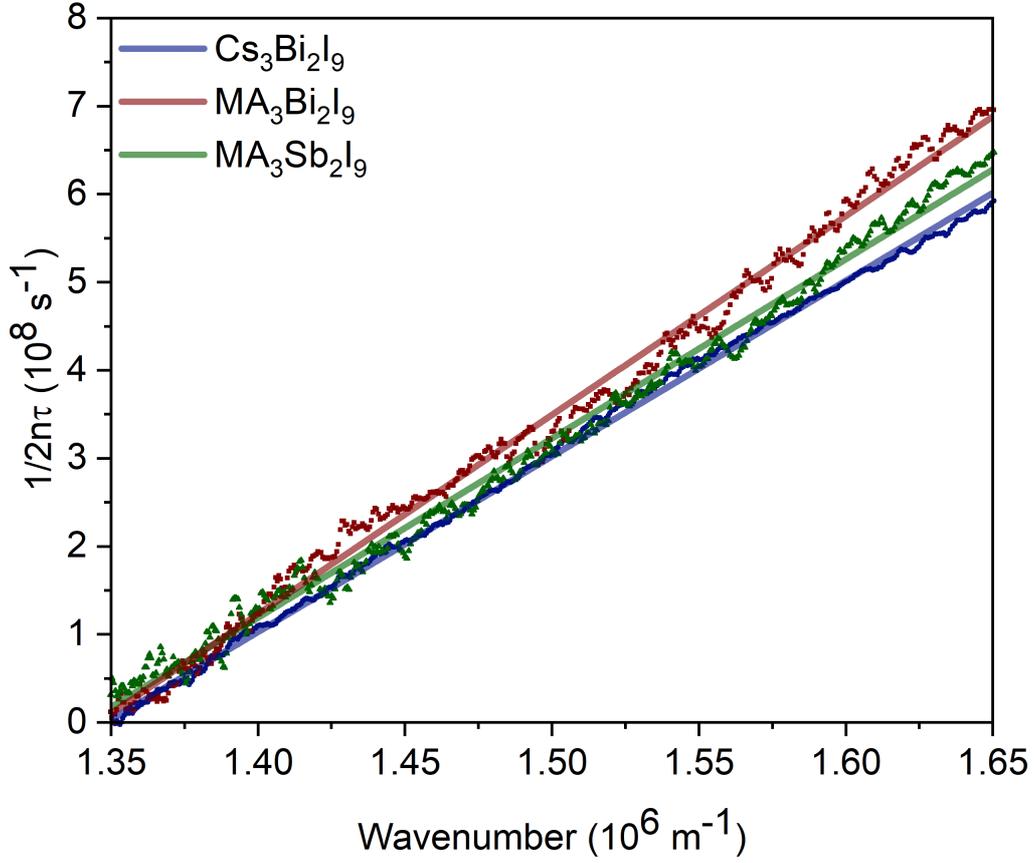

**Figure 4:** The oscillation frequency normalized by twice the refractive index ($1/2n\tau$) as a function of the probe wavenumber. The slope of the solid lines represents the longitudinal speed of sound ($V_L$). The curves are vertically shifted to underline the different slopes.

**Figure 4** shows the oscillation frequency normalized by twice the refractive index ($1/2n\tau$) as a function of the probe wavenumber. The refractive index dispersion is reported in the SM. The longitudinal speed of sound, $V_L$, is extracted by the slope of the curves of **Figure 4** for each material. In this case, eq. 1 becomes:

$$V_L = \sqrt{\frac{c_{11}}{\rho}} \qquad (4)$$

The extracted speeds of sound are reported in **Table 1**. The speed of sound measured for the inorganic $Cs_3Bi_2I_9$ agrees with values obtained with acousto-optic measurements [32], where the value for the longitudinal speed in the [001] direction was measured to be about 1980 m/s, and with previous TR measurements [24]. This agreement suggests that the results of our measurements in hybrid MA$_3$Sb$_2$I$_9$ and MA$_3$Bi$_2$I$_9$ represent a reliable evaluation of the speed of sound in these hybrid halide perovskites.

Once the speed of sound has been obtained, we evaluated the value of the $c_{11}$ elastic constant of the three materials using eq. 1. The results are reported in **Table 1**.

| Material | $V_L$ (m/s) | Density (Kg/m$^3$) | $c_{11}$ (GPa) |
|---|---|---|---|
| Cs$_3$Bi$_2$I | 1995±15 | 5230 | 20.8±0.2 |
| MA$_3$Bi$_2$I$_9$ | 2260±30 | 3010 | 15.4±0.2 |
| MA$_3$Sb$_2$I$_9$ | 2040±20 | 3610 | 15.0±0.2 |

**Table 1**. Longitudinal speed of sound ($V_L$), density, and $c_{11}$ elastic constant for the three materials investigated here.

As shown in **Table 1**, the $c_{11}$ elastic constant is higher in the inorganic material than in the hybrid compounds. This aspect is particularly evident from the comparison of Cs$_3$Bi$_2$I$_9$ and MA$_3$Bi$_2$I$_9$ that only differ for the monovalent A cation. On the contrary, $c_{11}$ is very similar in the two hybrid compounds where little effect is caused on the elastic properties by substituting Bi with Sb. A larger stiffness could justify a larger deformation potential of the energy bands and hence the larger amplitude of the oscillations shown in **Figure 3**.

In conclusion, we have measured the speed of sound in inorganic and hybrid halide perovskites, namely Cs$_3$Bi$_2$I$_9$, MA$_3$Bi$_2$I$_9$, and MA$_3$Sb$_2$I$_9$ using ultrafast transient optical reflectivity. The measurements have allowed the extraction of the speed of sound and the $c_{11}$ elastic constant in the single crystals of those materials. The value of $c_{11}$ in the inorganic Cs$_3$Bi$_2$I$_9$ is higher by about 30% than in the two hybrid materials thus indicating a lower stiffness for the latter, an aspect that has importance in designing optoelectronic devices.

## Acknowledgements

The authors from Brno University of Technology (BUT) acknowledge financial support from Grant Agency of the Czech Republic (GACR) project No. 25-17500S and Interfaculty project BUT No. FCH/FSI-J-24-8521. S.T., G.A., D.C., P.O.K., A.P., and F.T. acknowledge funding from the European Union – NextGenerationEU, M4C2, within the PNRR project NFFA-DI, CUP B53C22004310006, IR0000015.

The Supplemental Material file contains: Preparation of the samples and their microscopic images; Transient reflectivity maps for $Cs_3Bi_2I_9$ and (b) $MA_3Sb_2I_9$; Energy dispersion of the refractive index.

**Supplemental Materials**

**The speed of sound and elastic properties of single crystals of inorganic and hybrid lead-free iodide perovskites obtained by femtosecond transient optical reflectivity**

G. Ammirati[1], P. O'Keeffe[2], S. Turchini[1], D. Catone[1], A. Paladini[2], F. Toschi[1], S. Gavranovic[3], J. Posipil[3], G. Mannino[4], S. Valastro[4], and F. Martelli[1,5,*]

[1] *Istituto di Struttura della Materia - CNR (ISM-CNR), EuroFEL Support Laboratory (EFSL), Via del Fosso del Cavaliere 100, 00133, Rome, Italy.*
[2] *Istituto di Struttura della Materia - CNR (ISM-CNR), EuroFEL Support Laboratory (EFSL), Monterotondo Scalo 00015, Italy.*
[3]*Brno University of Technology Brno 612 00, Czech Republic*
[4]*IMM-CNR, Catania, Italy*
[5]*IMM-CNR, Rome, Italy*

*email: faustino.martelli@cnr.it

# S1. Preparation of halide perovskite single crystals

**Chemicals and reagents**

Cesium iodide (>99%, CAS No.: 14965-49-2), methylammonium iodide (≥99%, CAS No.: 14965-49-2), antimony (III) iodide (98%, CAS No.: 7790-44-5), hydroiodic acid (EMPLURA®, 57%, CAS No.: 10034-85-2), and N,N-dimethylformamide (99 % CAS No.: 68-12-2) were purchased from Sigma Aldrich (Merck). All chemicals were used without further purification.

**Hydrothermal synthesis**

The studied halide perovskite single crystals ($Cs_3Bi_2I_9$, $MA_3Bi_2I_9$, and $MA_3Sb_2I_9$) were synthesized using the hydrothermal (HT) method. Firstly, the 0,05 M perovskite solutions were prepared by dissolving CsI (MAI) and $BiI_3$ or $SbI_3$ (molar ratio 3:2) in 20 ml of hydroiodic acid (HI). The prepared solutions were then placed in the hydrothermal autoclave reactors, heated to 200 °C. In the next step, the perovskite solutions were slowly cooled down from 200 °C to 25 °C (with temperature gradient 1°C h$^{-1}$). The millimeter-sized single crystals were obtained after cooling process.

**Microscopic images of the prepared single crystals**

The images of halide perovskite single crystals presented below (Figures S1–S3) were captured using Hirox RH-2000 digital microscope under different magnification.

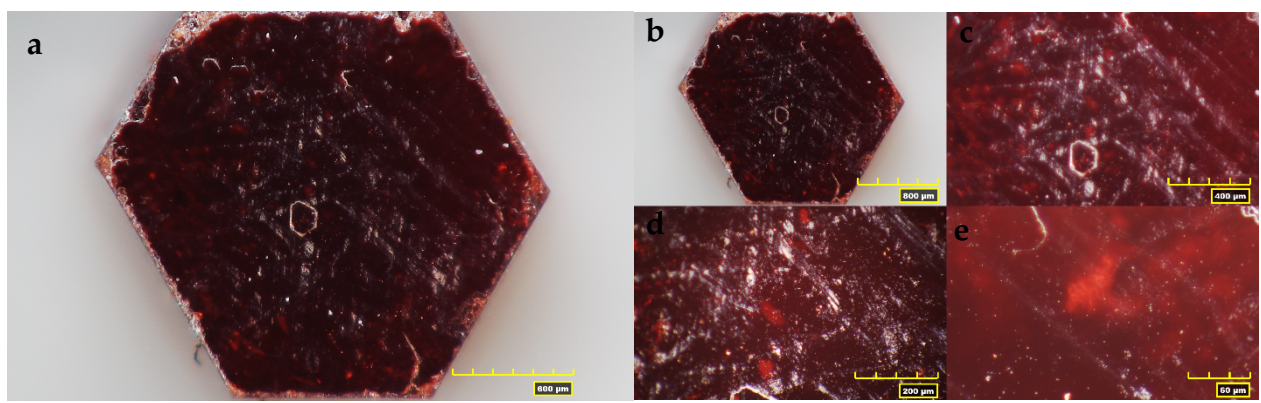

**Figure S1:** Microscopic image of the $Cs_3Bi_2I_9$ single crystal under different magnification (a) and (b) 50×, (c) 200×, (d) 400×, (e) 1 000×

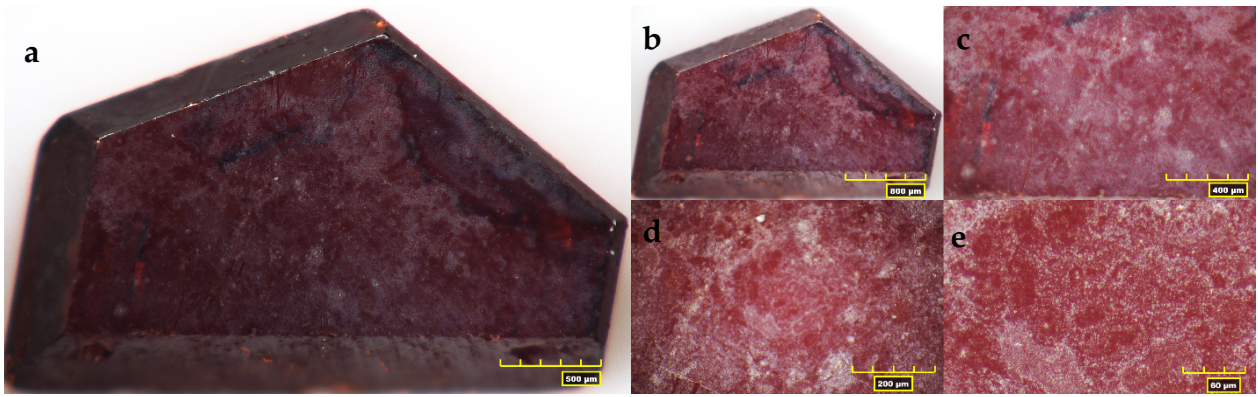

**Figure S2:** Microscopic image of the MA₃Bi₂I₉ single crystal under different magnification(a) and (b) 50×, (c) 200×, (d) 400×, (e) 1 000×.

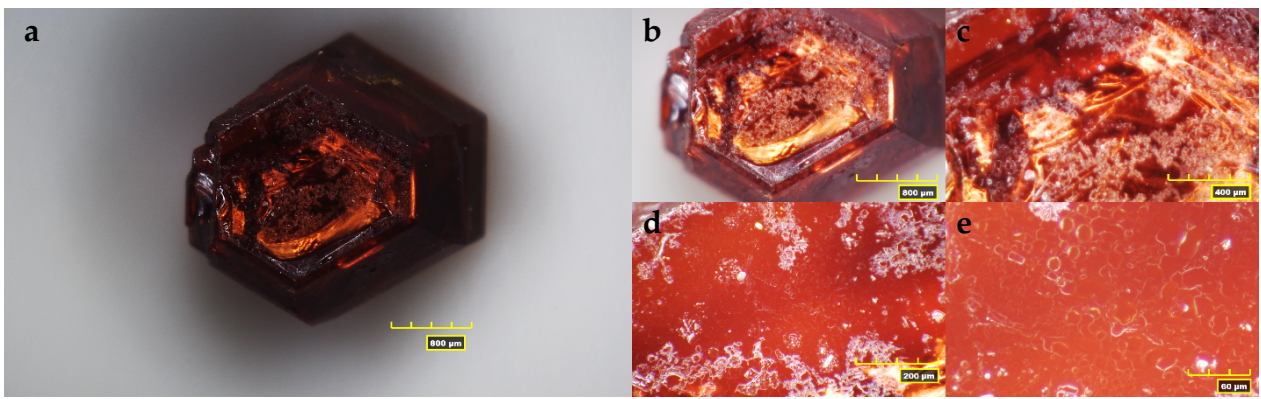

**Figure S3:** Microscopic image of the MA₃Bi₂I₉ single crystal under different magnification(a) 35×, (b) 50×, (c) 200×, (d) 400×, (e) 1 000×.

## S2. Transient Reflectivity Maps

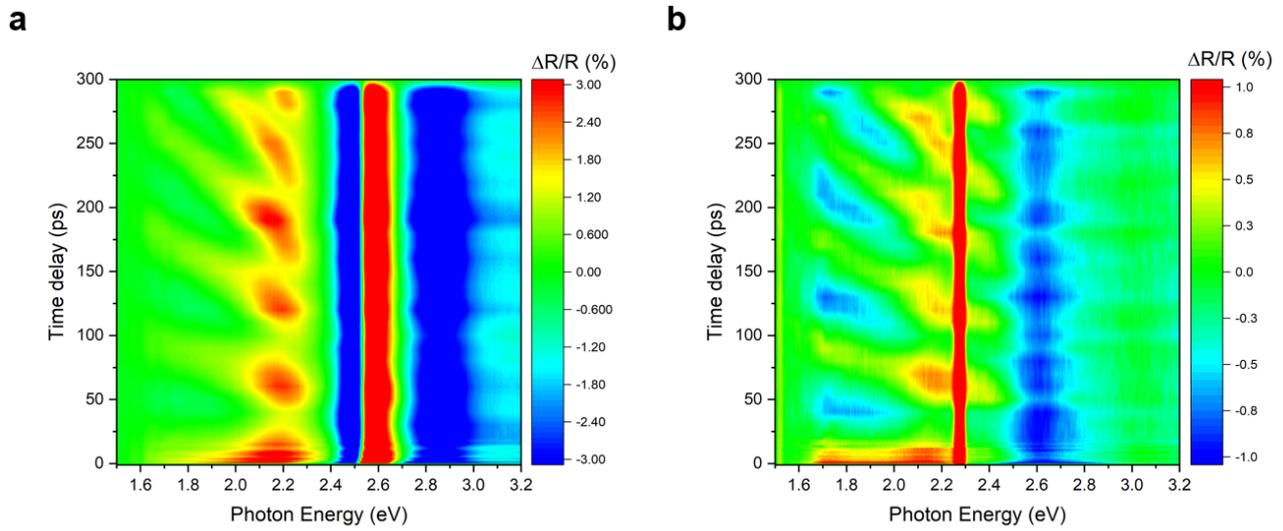

**Figure S4:** Transient reflectivity maps for (a) Cs$_3$Bi$_2$I$_9$ and (b) MA$_3$Sb$_2$I$_9$ obtained with pump energy of 4.50 eV and fluence of 600 µJ/cm$^2$.

# S3. Refractive Indexes obtained by spectroscopic ellipsometry

Spectroscopic Ellipsometry was performed by using a V-VASE, J.A. Woollam equipped with an autoretarder. The measurements have been performed at three angles, 50°, 60°, and 70° below and above the Brewster angle, over a wide range of wavelengths 245–1240 nm (1–4.5 eV) with steps of 10 nm or less depending on the curve steepness. Furthermore, as the sample was transparent in the sub-band gap spectral region, possible backside reflection was avoided by rear surface roughening via sandblasting. This strategy switches off any interference effect that affects optical measurements in the case of transparent substrates. A Kramers–Kronig consistent optical model was built based on multiple critical points parabolic band (CPPB) oscillators to fit experimental data ($\Psi$ and $\Delta$) and determine the real and imaginary parts of the dielectric function ($\varepsilon_1$ and $\varepsilon_2$). Measurements were collected using a slightly over-pressurized $N_2$-filled chamber to prevent sample degradation in the air.

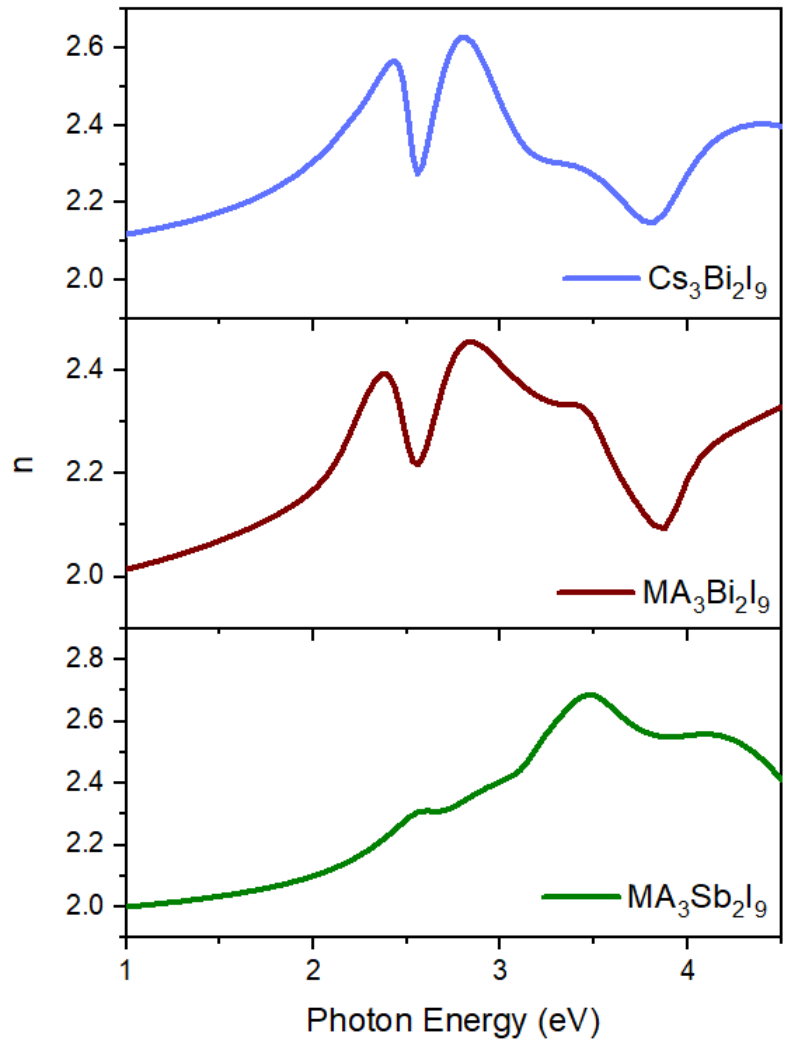

**Figure S5:** Refractive index of (a) Cs$_3$Bi$_2$I$_9$, (b) MA$_3$Bi$_2$I$_9$, and (c) MA$_3$Sb$_2$I$_9$ measured by spectroscopic ellipsometry.